\documentclass[usenatbib,fleqn,usedcolumn]{mnras}

\usepackage[british]{babel}
\usepackage{newtxtext}
\usepackage[slantedGreek]{newtxmath}

\let\la=\lesssim
\let\ga=\gtrsim

\usepackage[T1]{fontenc}
\usepackage{graphicx}

\usepackage{amsmath}	
\usepackage{amssymb}	

\usepackage{ctable}
\usepackage{url}
\usepackage{xspace}
\usepackage{pdflscape}

 \hypersetup{pdfauthor={C. C. Hayward et al.},
               pdftitle={Observational constraints on the physical nature of submillimetre source multiplicity},
               pdfkeywords={galaxies: high-redshift, galaxies: starburst, infrared: galaxies, submillimetre: galaxies},
               bookmarksnumbered=true}

\newcommand\sref[1]{\hyperref[#1]{Section~\ref*{#1}}}
\newcommand\fref[1]{\hyperref[#1]{Fig.~\ref*{#1}}}
\newcommand\Eqref[1]{equation~(\hyperref[#1]{\ref*{#1}})}
\newcommand\tref[1]{\hyperref[#1]{Table~\ref*{#1}}}

\newcommand\msunperyr{\text{M}_{\sun}~\text{yr}^{-1}\xspace}
\newcommand\zphot{z_{\text{phot}}}
\newcommand\zspec{z_{\text{spec}}}

\newcommand{\nodata}{\ldots}

\title[Physical nature of submm source multiplicity]{Observational constraints on the physical nature of submillimetre source multiplicity:~chance projections are common}
\author[Hayward, Chapman, Steidel et al.]{
\parbox[t]{\textwidth}{
Christopher C.~Hayward$^{1,2,3}$\thanks{E-mail: \href{mailto:chayward@flatironinstitute.org}{chayward@flatironinstitute.org}}, Scott C.~Chapman$^4$, Charles C.~Steidel$^3$, Anneya Golob$^5$,
Caitlin M.~Casey$^6$, Daniel J.~B. Smith$^7$, Adi Zitrin$^8$, Andrew W.~Blain$^9$, Malcolm N.~Bremer$^{10}$, Chian-Chou Chen$^{11}$, Kristen E.~K.~Coppin$^{7}$, Duncan Farrah$^{12}$,
Eduardo Ibar$^{13}$, Micha{\l} J.~Micha{\l}owski$^{14}$, Marcin Sawicki$^5$, Douglas Scott$^{15}$,
Paul van der Werf$^{16}$, Giovanni G.~Fazio$^2$, James E.~Geach$^7$, Mark Gurwell$^2$, Glen Petitpas$^2$, and David J.~Wilner$^2$
}
\vspace*{6pt} \\
$^1$Center for Computational Astrophysics, Flatiron Institute, 162 Fifth Avenue, New York, NY 10010, USA\\
$^2$Harvard--Smithsonian Center for Astrophysics, 60 Garden Street, Cambridge, MA 02138, USA\\
$^3$California Institute of Technology, 1200 E. California Boulevard, Pasadena, CA 91125, USA\\
$^4$Department of Physics and Atmospheric Science, Dalhousie University, 6310 Coburg Road, Halifax, NS B3H 4R2, Canada\\
$^5$Department of Astronomy and Physics and the Institute for Computational Astrophysics, Saint Mary's University, 923 Robie Street, Halifax, Nova Scotia, \\ B3H 3C3, Canada\\
$^6$Department of Astronomy, The University of Texas at Austin, 2515 Speedway Boulevard Stop C1400, Austin, TX 78712, USA\\
$^7$Centre for Astrophysics Research, University of Hertfordshire, College Lane, Hatfield, Hertfordshire, AL10 9AB, UK\\
$^8$Physics Department, Ben-Gurion University of the Negev, P.O. Box 653, Be'er-Sheva 8410501, Israel\\
$^9$Physics \& Astronomy, University of Leicester, 1 University Road, Leicester LE1 7RH, UK\\
$^{10}$H.~H. Wills Physics Laboratory, University of Bristol, Tyndall Avenue, Bristol, BS8 1TL, UK\\
$^{11}$European Southern Observatory, Karl Schwarzschild Stra{\ss}e 2, 85748 Garching, Germany\\
$^{12}$Department of Physics, Virginia Tech, Blacksburg, VA 24061, USA\\
$^{13}$Instituto de F\'isica y Astronom\'ia, Universidad de Valpara\'iso, Avda. Gran Breta\~na 1111, Valpara\'iso, Chile\\
$^{14}$Astronomical Observatory Institute, Faculty of Physics, Adam Mickiewicz University, ul.~S{\l}oneczna 36, 60-286 Pozna{\'n}, Poland\\
$^{15}$Department of Physics and Astronomy, University of British Columbia, 6224 Agricultural Road, Vancouver, BC V6T 1Z1, Canada\\
$^{16}$Leiden Observatory, Leiden University, P.O. Box 9513, NL-2300 RA Leiden, The Netherlands
}

\date{MNRAS, in press}
\pubyear{2018}

\begin{document}
\label{firstpage}
\pagerange{\pageref{firstpage}--\pageref{lastpage}}
\maketitle

\interfootnotelinepenalty=10000

\begin{abstract}
Interferometric observations have demonstrated that a significant fraction of single-dish submillimetre (submm) sources
are blends of multiple submm galaxies (SMGs), but the nature of this multiplicity, i.e. whether the galaxies are physically associated or chance projections,
has not been determined.
We performed spectroscopy of 11 SMGs in six multi-component submm sources, obtaining spectroscopic redshifts for nine of them.
For an additional two component SMGs, we detected continuum emission but no obvious features. We supplement our observed sources with four single-dish submm sources from
the literature. This sample allows us to statistically constrain the physical nature of single-dish submm source multiplicity for the first time.
In three $\left(3/7, \text{ or } 43\substack{+39 \\ -33} \text{ per cent at 95\% confidence} \right)$ of the single-dish sources for which the nature of the blending is unambiguous,
the components for which spectroscopic redshifts are available are physically associated, whereas 4/7 $\left(57\substack{+33 \\ -39} \text{ per cent} \right)$
have at least one unassociated component. When components whose spectra exhibit continuum but no features
and for which the photometric redshift is significantly different from the spectroscopic redshift of the other component are also considered,
6/9 $\left(67\substack{+26 \\ -37} \text{ per cent}\right)$ of the single-dish sources are comprised of at least one unassociated component SMG.
The nature of the multiplicity of one single-dish source is ambiguous.
We conclude that physically associated systems and chance projections both contribute to the multi-component single-dish submm source
population. This result contradicts the conventional wisdom that bright submm sources are solely a result of merger-induced starbursts, as blending of unassociated
galaxies is also important.
\end{abstract}


\section{Introduction} \label{S:intro}

Submillimetre (submm) wavelengths are ideal for selecting galaxies across a wide range of redshift because of the so-called
`negative $k$-correction' \citep[e.g.][]{Blain:2002}, but their drawback is that the beam sizes of typical single-dish submm telescopes are $\ga 15$ arcsec
($\ga 120$ kpc at $z \ga 1$).\footnote{Throughout this work, we assume $\Omega_{\text{m}} = 0.31$, $\Omega_{\Lambdaup} = 0.69$, and $H_0 = 68$
km s$^{-1}$ Mpc$^{-1}$ \citep{Planck2015}.} Consequently, blending of emission from more than one galaxy into one single-dish submm source is possible.
Such blending is more likely for submm sources than for e.g. optical sources because of both the large beam size and the negative $k$-correction, which jointly imply,
roughly speaking, that a galaxy with a fixed spectral energy distribution will contribute approximately the same flux to the observed single-dish submm source if the galaxy is located anywhere within
a cylinder of diameter $\sim 240$ kpc spanning $z \sim 1-10$. This potential problem has long been recognised \citep[e.g.][]{Hughes:1998}, but constraining
the prevalence and nature of blended submm sources has been challenging. The fact that single-dish submm sources often have multiple radio
\citep[e.g.][]{Ivison:2002,Ivison:2007} and $K$-band \citep{Smith2017} counterparts suggests that blending is common. However, to directly determine whether single-dish submm sources are
actually blends of multiple sources, interferometric observations of the dust continuum emission are required.

Before the Atacama Large Millimeter/submillimeter Array (ALMA\footnote{\url{http://www.almaobservatory.org/}})
came online, only a handful of single-dish submm sources
were observed with submm interferometers such as the Submillimeter Array (SMA\footnote{\url{https://www.cfa.harvard.edu/sma/}})
and the Plateau de Bure Interferometer (PdBI\footnote{\url{www.iram-institute.org/EN/plateau-de-bure.php}});
some of these pre-ALMA interferometric studies resolved single-dish submm sources into multiple components
\citep[e.g.][]{Younger:2009SMG_interf,Engel:2010,Wang:2011,Barger:2012,Smolcic:2012}.
In the era of ALMA, it has become possible to interferometrically map large numbers of single-dish submm sources.
To date, ALMA follow-up observations have demonstrated
that many single-dish submm sources are blends of two or more resolved SMGs
(e.g. \citealt{Karim2013,Hodge2013,Wiklind:2014,Bussmann2015,Simpson2015b,Simpson2015}; see \citealt{Casey2014} for a review),
although details such as the fraction of single-dish submm sources that are blends and the effect of blending
on submm number counts are still debated \citep[e.g.][]{Chen2013,Koprowski:2014,Michalowski2017}.

Although it is clear that a significant fraction of single-dish submm sources are blends, the physical nature of this multiplicity
has not been constrained. Specifically, are the individual components of multi-component single-dish submm sources
physically associated galaxies, either undergoing a merger or within the same dark matter structure (e.g. group or filament) but
not actively merging? Or are blended submm sources chance projections of
galaxies at different redshifts that have no dynamical influence on one another?

Various theoretical works have considered the effects of blending on submm number counts. \citet{H11} were the first to
suggest that pre-coalescence galaxy mergers (i.e.~blends of two physically associated SMGs) should contribute significantly
to the single-dish submm number counts. In follow-up work, \citet{H12} discussed how to distinguish pre-coalescence mergers
(`galaxy-pair SMGs') from coalescence-stage starbursts, and \citet{HN13} presented detailed predictions for the relative
contributions of merger-induced starbursts, blended pre-coalescence merging galaxies, and isolated discs to
the submm number counts. In the model of \citet{Narayanan2015}, the contribution of satellite galaxies
results in blended sources comprised of physically associated component SMGs. \citet{HB13} were the first to investigate the effects of blending of \emph{unassociated}
galaxies; they predicted that of the subset of single-dish submm sources that are blends of multiple SMGs, the majority should be comprised
of at least one SMG that is physically unassociated with the other component(s). Using a semi-analytic model in which the
physical nature of SMGs is drastically different than in the \citet{HB13} model, \citet{Cowley2015a} concluded that effectively
all multi-component single-dish submm sources should be comprised of at least one unassociated SMG.
Applying a `counts-matching' approach to a semi-analytic model, \citet{MunozArancibia2015} also predicted that the majority
of the components of blended single-dish submm sources are spatially unassociated.
Notably, all theoretical models that treat blending of unassociated galaxies predict that chance projections contribute significantly
to the multiple-component single-dish submm source population, but these predictions have not been tested to date.

It is possible to indirectly constrain the relative importance of the two types of blending using the distribution of
the angular separations of the resolved submm components of single-dish submm sources, but such analyses
have yielded conflicting results \citep[see the comparison presented in fig.~7 of][]{Bussmann2015}, and the
indirect nature of such constraints makes redshift-based constraints preferable.
Interferometric maps of the dust continuum emission are a prerequisite for obtaining accurate redshifts because relying
on potential counterparts at other wavelengths runs the risk of yielding a redshift for a galaxy that does not
actually contribute to the single-dish submm source.
Even when the positions of the resolved submm components are known,
obtaining redshifts is challenging because photometric redshifts of SMGs
may not be sufficiently accurate to constrain the physical nature of the blending \citep{Simpson2014}.
Spectroscopic redshifts can provide unambiguous constrains, but obtaining spectroscopic redshifts for SMGs is notoriously difficult for
multiple reasons. For example, at fixed submm flux density, the optical/near-infrared (NIR) emission line luminosities
of individual SMGs can differ considerably, probably owing to patchy dust extinction.
Consequently, optical/NIR spectroscopic follow-up often yields non-detections \citep[e.g.][]{Danielson2017}. Obtaining redshifts via molecular and atomic
emission lines in the far-infrared (FIR) and submm is an alternative approach, and even unresolved observations can reveal chance projections \citep{Zavala2015}.
However, the overhead associated with multiple tunings makes this observationally expensive, and
often, guided by photometric redshifts, only a limited wavelength -- and thus redshift
-- range is probed.\footnote{Given the bias against chance projections inherent in these observations, in this work, we do not consider
FIR/submm atomic or molecular gas emission line-based redshifts from the literature except for COSBO3, for which we
complement H$\alphaup$-based redshifts with some CO-based redshifts. Excluding the CO-based redshifts does not qualitatively
affect our conclusions.}
For these reasons, very few multi-component single-dish submm sources for which spectroscopic redshifts of at least two component SMGs are available
have been presented in the literature \citep{Wang:2011,Barger:2012,Danielson2017}.

We present spectroscopic constraints on the redshifts of components of six single-dish submm sources based on spectra obtained with
Keck\footnote{\url{http://www.keckobservatory.org/}}, Gemini\footnote{\url{http://www.gemini.edu/}} and
the Very Large Telescope\footnote{\url{http://www.eso.org/public/usa/teles-instr/paranal-observatory/vlt/}}.
We supplement our observational dataset with four single-dish submm sources from the literature. The combined sample
is comprised of seven multi-component single-dish submm sources for which optical emission line-based spectroscopic redshifts are available for two or more
components. For an additional three single-dish sources, a spectroscopic redshift is available for one component, and continuum only is detected in the
spectrum of the other; for these sources, photometric redshift estimates for the continuum-only components are available, and we use these to tentatively
constrain the nature of the multiplicity of these sources. For the first time, we are able to constrain the relative contributions of physically
associated galaxies and chance projections to the multi-component single-dish submm source population.

\section{Observational data} \label{S:methods}

\ctable[
	caption = {Properties of the components of the single-dish submm sources in our sample. \label{tab:sample_details}},
	center,
	doinside=\scriptsize,
	star,
	notespar
]{lccccclccccc}{
	\tnote[a]{Peak SMA 860 \micron\xspace \citep{Hill2017} or ALMA 870 \micron\xspace \citep{Simpson2015} flux density.}
	\tnote[b]{Source of submm interferometric observations: (1) SMA \citep{Hill2017}; (2) ALMA \citep{Simpson2015,Simpson2015b}; (3) ALMA \citep{Bussmann2015}; (4) SMA \citep{Wang:2011};
		     (5) SMA \citep{Barger:2012}; (6) ALMA \citep{Hodge2013}.}
	\tnote[c]{Reference for $\zspec$ ($\zphot$ reference in parentheses) or spectral follow-up if undetected or continuum-only detection: (7) \citet{Strazzullo2010}; (8) \citet{Roseboom2012};
			(9) \citet{Simpson2017}; (10) \citet{Wang2016}; (11) Champagne et al., in prep.; (12) \citet{Laigle2016}; 
			(13) \citet{Barger:2008}; (14) \citet{Trouille2008}; (15) \citet{Chapman:2005}; (16) \citet{Bluck2012}; (17) \citet{Wang2006}; (18) \citet{Swinbank2004}; (19) \citet{Barger2014};
			(20) \citet{Danielson2017} -- the Q values represent the spectral quality, with $Q = 3$ indicating redshifts that should not be considered fully independent of the photometric redshifts and $Q = 4$
			corresponding spectra that exhibit continuum but no sufficiently trustworthy features to assign a spectroscopic redshift; see \citet{Danielson2017} for further details; (21) \citet{Simpson2014}.
			For spectroscopic redshifts obtained in this work, the instrument is denoted by a letter in brackets: M = MOSFIRE, G = GNIRS, X = XSHOOTER.}
	\tnote[$\ast$]{The errors on the spectroscopic redshifts reported in this work are $dz \sim 0.0005$.}
	\tnote[\dag]{Spectrum exhibits continuum only; thus, the spectroscopic redshift is unconstrained.}
	\tnote[$\star$]{The NIR source is offset by 2.5 arcsec ($\sim 20$ kpc at $z = 2.6$) from the ALMA position. It is thus likely that the NIR and ALMA sources are unrelated, and the galaxy
	responsible for the submm emission may not be at $z = 2.606$.}
	}{
													\FL
Single-dish ID	&	Component ID	&	RA (J2000.0)	&	Dec (J2000.0)		&	$S_{860/870}$\tmark[a]	&	$\zspec$\tmark[$\ast$]	&	$\zphot$	&	Lines detected	&	Submm ref\tmark[b] 	&	$z$ ref\tmark[c]
&	Figure \NN
			&			& (h:m:s)		& (d:m:s)			& 	(mJy)				&		&						&							&		&					& \ML
LOCK-03		&	a		& 10:47:27.97	& $+$58:52:14.1	&	$8.1 \pm 1.8$			& 2.209	& $2.76 \pm 0.18$			& H$\alphaup$, [\ion{N}{ii}]		&	1	&	This work [M] (7)	& \ref{fig:mosfire_spectra} \NN 
\noalign{\smallskip}
			&	b		& 10:47:26.52	& $+$58:52:12.8	&	$8.0 \pm 1.9$			& 2.363	& $2.10 \pm 0.34$			& H$\alphaup$, [\ion{N}{ii}]		&	1	&	This work [M] (7)	& \ref{fig:mosfire_spectra} \NN 
\noalign{\smallskip}
\noalign{\smallskip}
LOCK-08		&	a		& 10:47:00.18	& $+$59:01:07.5	&	$10.4 \pm 1.6$			& 2.279	& 2.06$^{+0.09}_{-0.18}$		& H$\alphaup$					&	1	&	This work [M] (8)	& \ref{fig:mosfire_spectra} \NN
\noalign{\smallskip}
			&	b		& 10:47:02.48	& $+$59:00:50.3	&	$4.8 \pm 1.6$			& 2.280	& \nodata					& H$\alphaup$, [\ion{N}{ii}]		&	1	&	This work [M] 		& \ref{fig:mosfire_spectra} \NN 
\noalign{\smallskip}
\noalign{\smallskip}
LOCK-09		&	a		& 10:45:23.11	& $+$59:16:18.6	&	$9.4 \pm 1.5$			& \nodata	& \nodata					& \nodata						&	1	&					& \NN
\noalign{\smallskip}
			&	b		& 10:45:24.94	& $+$59:16:26.7	&	$5.1 \pm 1.5$			& 1.633	& $1.46 \pm 0.19$			& H$\alphaup$, [\ion{N}{ii}]		&	1	&	This work [M] (7)	& \ref{fig:lh09_spectra} \NN 
\noalign{\smallskip}
			&	c		& 10:45:23.55	& $+$59:16:32.2	&	$4.5 \pm 1.5$			& \dag 	& $0.90 \pm 0.05$			& continuum					&	1	&	This work [M] (7)	& \ref{fig:lh09_spectra} \NN
\noalign{\smallskip}
\noalign{\smallskip}
UDS292		&	0		& 2:17:21.53	& $-$5:19:07.8		&	$4.2 \pm 0.8$			& 2.383	& $2.65^{+0.25}_{-0.07}$		& [\ion{O}{iii}]4959, 5007			&	2	&	This work [G] (9) 	& \ref{fig:gnirs_spectra} \NN
\noalign{\smallskip}
			&	1		& 2:17:21.96	& $-$5:19:09.8		& 	$3.9 \pm 0.8$			& 2.387	& $2.51^{+0.23}_{-0.10}$		& H$\alphaup$					&	2	&	This work [G] (9) 	& \ref{fig:gnirs_spectra} \NN
\noalign{\smallskip}
\noalign{\smallskip}
UDS306		&	0		& 2:17:17.07	& $-$5:33:26.6		&	$8.3 \pm 0.5$			& 2.603	& $2.31^{+0.06}_{-0.21}$		& H$\alphaup$, [\ion{O}{ii}]		&	2	&	This work	[G] (9) 	& \ref{fig:gnirs_spectra} \NN
\noalign{\smallskip}
	 		&	1		& 2:17:17.16	& $-$5:33:32.5		&	$2.6	\pm 0.4$			& \dag	& $1.28^{+0.53}_{-0.06}$		& continuum					&	2	&	This work	 [G] (9) 	& \ref{fig:gnirs_spectra} \NN
\noalign{\smallskip}
			&	2		& 2:17:16.81	& $-$5:33:31.8		&	$3.0 \pm 0.9$			& 2.606$^{\star}$	& \nodata			& [\ion{O}{iii}]4959, 5007			&	2	&	This work [X]		& \ref{fig:xshooter_spectra} \NN
\noalign{\smallskip}
\noalign{\smallskip}
COSBO3		&	a		& 10:00:56.95	& $+$2:20:17.3		& 	$5.3 \pm 0.3$			& 2.494	& \nodata					& CO(1-0), CO(3-2)				& 	3	&	10, 11					\NN
\noalign{\smallskip}
			&	b		& 10:00:57.57	& $+$2:20:11.2		&	$3.8 \pm 0.3$			& 2.513	& $2.71^{+0.15}_{-0.13}$		& H$\alphaup$, CO(1-0)			&	3	&	10, 11 (12)				\NN
\noalign{\smallskip}
			& 	c		& 10:00:57.27	& $+$2:20:12.7		&	$1.7 \pm 0.2$			& 2.498	& $2.18^{+0.10}_{-0.08}$		& H$\alphaup$					&	3	&	This work [M] (12)	& \ref{fig:mosfire_spectra} \NN
\noalign{\smallskip}
			&	d		& 10:00:57.40	& $+$2:20:10.8		&	$2.2 \pm 0.4$			& 2.508	& $2.31^{+0.03}_{-0.04}$		& CO(1-0)						&	3	&	10 (12)					\NN
\noalign{\smallskip}
			&	e		& 10:00:56.86	& $+$2:20:08.9		&	$1.69 \pm 0.3$			& 2.503	& $2.28^{+0.05}_{-0.06}$		& CO(1-0)						&	3	&	10 (12)					\NN
\noalign{\smallskip}
\noalign{\smallskip}
GOODS 850-13 &	a		& 12:37:14.03	& $+$62:11:56.4	&	$3.2 \pm 0.9$			& \nodata	& 3.46$^{+1.04}_{-0.86}$		& \nodata						&	4	&	(3)			\NN
\noalign{\smallskip}
			&	b		& 12:37:14.26	& $+$62:12:08.1	&	$4.1 \pm 0.7$			& 3.157	& 1.25					& not reported					&	4	&	17 (18)		\NN
\noalign{\smallskip}
			&	c		& 12:37:12.00	& $+$62:12:12.3	&	$5.3 \pm 0.9$			& 2.914	& 2.80					& Ly$\alphaup$, \ion{C}{iv}		&	4	&	19 (16)		\NN
\noalign{\smallskip}
\noalign{\smallskip}
GOODS 850-15 &	a		& 12:36:21.10	& $+$62:17:09.6	&	$3.4 \pm 0.6$			& 1.988	& 2.91					& H$\alphaup$, interstellar abs. lines	&	5	&	19 (17)		\NN
\noalign{\smallskip}
			&	b		& 12:36:21.30	& $+$62:17:08.1	&	$3.5 \pm 0.7$			& 1.992	& 2.016					& H$\alphaup$, [\ion{N}{ii}]		&	5	&	18 (19)		\NN
\noalign{\smallskip}
\noalign{\smallskip}
ALESS 084	&	1		& 3:31:54.50	& $-$27:51:05.6 	&	$3.2 \pm 0.6$			& 3.965	& 	1.92$^{+0.09}_{-0.07}$	& Ly$\alphaup$, \ion{N}{v}, cont. ($Q=3$)		&	6	&	20 (21)	\NN 
\noalign{\smallskip}
			&	2		& 3:31:53.85	& $-$27:51:04.4 	&	$3.2 \pm 0.8$			& \dag 	& 	1.75$^{+0.08}_{-0.19}$	& cont., poss. faint lines ($Q=4$)			&	6	&	20 (21)	\NN
\noalign{\smallskip}
\noalign{\smallskip}
ALESS 088	&	1		& 3:31:54.76	& $-$27:53:41.5 	&	$4.6 \pm 0.6$			& 1.268	& 	1.84$^{+0.12}_{-0.11}$	& [\ion{O}{ii}], [\ion{O}{ii}]3726,3729 ($Q=1$)	&	6	&	20 (21)	\NN 
\noalign{\smallskip}
			&	2		& 3:31:55.39	& $-$27:53:40.3 	&	$2.1 \pm 0.5$			& 2.519	& 	\nodata 				& \ion{C}{ii}]2326, \ion{C}{iv} ($Q=3$)		&	6	&	20	\NN 
\noalign{\smallskip}
			&	5		& 3:31:55.81	& $-$27:53:47.2 	&	$2.9 \pm 0.7$			& 2.294	& 	2.30$^{+0.11}_{-0.50}$	& Ly$\alphaup$, \ion{He}{ii}, cont. ($Q=2$)	&	6	&	20 (21)	\NN 
\noalign{\smallskip}
			&	11		& 3:31:54.95	& $-$27:53:37.6 	&	$2.5 \pm 0.7$			& 2.358	& 	2.57$^{+0.04}_{-0.12}$	& Ly$\alphaup$, \ion{C}{iii}] ($Q=3$)			&	6	&	20 (21)	\LL 
}

\subsection{S2CLS sources observed in this study}

Details regarding our single-dish submm source sample, including both those observed in this work and those drawn from the literature,
are presented in \tref{tab:sample_details}.
The single-dish submm sources observed as part of this study were primarily drawn from the Submillimetre Common-User Bolometer Array 2
(SCUBA-2\footnote{\url{http://www.eaobservatory.org/jcmt/instrumentation/continuum/scuba-2/}}) Cosmology Legacy Survey at 850 \micron\xspace
\citep[S2CLS\footnote{\url{http://www.astro.dur.ac.uk/~irs/S2CLS/}};][]{Geach2017}.
\citet{Simpson2015b,Simpson2015} followed-up 30 bright ($S_{870} \ga 8$ mJy) S2CLS single-dish submm sources with ALMA. An additional
75 single-dish submm sources with $S_{870} > 8$ mJy were followed-up with the SMA;
the catalogue and SMA observations are detailed in \citet{Hill2017}.
The interferometric observations reveal whether the S2CLS single-dish submm sources are composed of multiple SMGs.
We selected the brightest multiple-component single-dish submm sources from the S2CLS for which SMA or ALMA follow-up observations were
available prior to our observing nights as candidates for NIR spectroscopic follow-up.
To avoid selecting an incorrect counterpart while simultaneously minimizing the chance of following-up
spurious ALMA or SMA sources, we targeted multiples for which at least one -- and preferably more than one --
of the ALMA or SMA sources had an
unambiguous companion at optical or NIR wavelengths. Note that this selection may make our
sample biased toward lower-redshift or/and more-massive SMGs and against chance projections in which the secondary submm component is located at a
significantly greater redshift than the primary and thus may not be visible in the rest-frame optical.

\subsubsection{MOSFIRE observations}

We observed two component SMGs of each of two
multi-component single-dish SCUBA-2 850-\micron\xspace sources (LOCK-03 and LOCK-09) using the Multi-Object Spectrometer for InfraRed Exploration
\citep[MOSFIRE\footnote{\url{https://www2.keck.hawaii.edu/inst/mosfire/}};][]{McLean2010,McLean2012}, an NIR imaging spectrometer on the Keck 1 telescope,
on 25 February 2016. We additionally observed the two component SMGs of LOCK-08; because these are separated by 25 arcsec, they are not blended in
the SCUBA-2 map, but they would be blended in e.g. LABOCA 870-\micron\xspace or SPIRE 350- and 500-\micron\xspace maps.
We detected line or/and continuum from all six of the resolved SMGs.
The slitmasks were designed to also include some single-component submm sources, only one component of
some multi-component submm sources, and any nearby radio sources that could be accommodated within the
$3^{\arcmin} \times 6.1^{\arcmin}$ field of view.

All slits were 0.7 arcsec in width, resulting in spectra with resolving power $R \approx 3650$ in
the \emph{H}- or \emph{K}-band atmospheric windows.
Given that the typical sizes of SMGs ($R_\mathrm{e} \sim 1$ kpc, as determined from submm emission; e.g. \citealt{Simpson2015}) are much smaller
than the slit width (approximately 6 kpc at the redshifts of interest) and that the seeing was good (0.65 arcsec),
slit losses are likely $\lesssim 30$ per cent. Moreover, because we are primarily concerned with redshifts rather than line luminosities,
aperture corrections to the line fluxes are unnecessary.

The spectra were obtained using the standard two-position `mask nod' in which the telescope position was dithered
$\pm 1.5$ arcsec along the slit direction, with individual integration times of 120 s (\emph{H}-band) or 180 s (\emph{K}-band). 
Total integration times of 2880 s were obtained, with 24 and 16 individual exposures in the \emph{H} and \emph{K} bands, respectively. 

The MOSFIRE data were reduced using the publicly available data reduction pipeline developed by the instrument
team;\footnote{\url{https://keck-datareductionpipelines.github.io/MosfireDRP/}.} see \citet{Steidel:2014} for a detailed description of the procedure. 
One-dimensional spectra were extracted, flux-calibrated, and analysed using the MOSPEC package \citep{Strom2017}.

\subsubsection{GNIRS observations}
\label{sec:GNIRS}

Near-infrared spectra of resolved components of two multi-component single-dish submm sources (UDS292 and UDS306)
were obtained using the cross-dispersed mode of the Gemini Near-Infrared Spectrograph (GNIRS\footnote{\url{http://www.gemini.edu/sciops/instruments/gnirs/}})
on the Gemini North 8.1-m telescope.
This configuration provides continuous spectral coverage from $0.84 - 2.48$ \micron\xspace at a spectral resolution of $R\approx1500$ with a spatial scale of 0.15 arcsec/pixel.
The slit dimensions were 1.0\arcsec ~$\times$ 7.0\arcsec.
The observations used an ABBA pattern, nodding along the slit to keep the galaxy on slit at all times. Eight individual on-source integrations of 240 s each were performed for each source.
Observations of standard stars were obtained before and after each set of SMG observations. These were used to correct the spectra for telluric absorption.

The spectral reduction, extraction, and wavelength and flux calibration procedures were performed using the Gemini IRAF package and PyRAF.\footnote{PyRAF is a product of the Space
Telescope Science Institute, which is operated by AURA for NASA.} Briefly, the processing consists of removing cosmic ray-like features, dividing by flat fields taken with infrared and
quartz halogen lamps, subtracting sky emission using exposures taken at a different point in the dither pattern, and finally rectifying the tilted, curved spectra using pinhole flats.
Wavelength calibration is performed using Argon arc spectra, and then a spectrum of each order is extracted, divided by the standard star observation to cancel telluric absorption lines, and
roughly flux-calibrated using the telluric standard star spectrum. The different spectral orders for each extraction window are merged into a single 1D spectrum from $0.84-2.48$ \micron\xspace.
In all cases, the agreement in flux between the overlapping regions of two consecutive orders was very good, and scaling factors of only 3 per cent or less were necessary.

\subsubsection{XSHOOTER observations}

One target was observed on 4 March 2015 with the XSHOOTER echelon spectrograph \citep{Vernet2011} on
the ESO VLT-UT2 (Kueyen) as part of programme 094.A-0811(A), providing near-continuous spectroscopy from $0.3-2.48$ \micron\xspace with a 1.2\arcsec-wide and 11\arcsec-long slit.
The slit was located on a possible NIR counterpart (but see footnote \ref{foot306}) located 2.5 arcsec from the ALMA source UDS 306.2,
dithering the observations in an ABBA sequence at positions $+3$ and $-3$ arcsec along the slit axis.
The observation was setup to first peak up on a nearby star in a field within 1 arcmin of the target position, and then a blind offset was performed.
Twelve exposures of 300 seconds each were taken.

The ESO XSHOOTER pipeline \citep{Modigliani2010} was used to reduce the data. This pipeline was used to perform spatial and spectral rectification on the spectra (which exhibited significant spatial curvature
in addition to a non-linear wavelength scale) by using two-dimensional arc spectra obtained through a pinhole mask. For the IR channel, the data were mapped to an output spectral scale
of 1\,\AA\,pix$^{-1}$ and a spatial scale of 0.21 arcsec (from original scales of approximately 0.5\,\AA\,pix$^{-1}$ and 0.24 arcsec, respectively).
In the optical channel, the data were mapped to an output spectral scale of 0.4\,\AA\,pix$^{-1}$ and a spatial scale of 0.16 arcsec.
In both channels, the data were flat-fielded, and cosmic rays were identified and masked.
The two dither positions were subtracted to remove the sky to first order, and the different echelle orders were combined together into a continuous spectrum (taking into account
the variation in throughput with wavelength in different overlapping echelle orders) before spatially registering and combining the data taken at the two dither positions and
removing any residual sky background.

\subsection{COSBO3}

We also include COSBO3 (aka AzTEC-C6 and COSMOS 850.05), for which we obtained an H$\alphaup$-based redshift for one component SMG in this work; H$\alpha$- and
CO-based redshifts for additional component SMGs have been presented in the literature previously.
COSBO3 was originally detected as a single millimetre source by MAMBO \citep{Bertoldi2007} and AzTEC \citep{Aretxaga2011} and at 850 \micron\xspace with SCUBA-2 \citep{Casey2013}.
The accompanying 450-\micron\xspace maps from SCUBA-2 hinted at multiplicity, with the source being split into two independent sources at 7-arcsec resolution (identified in \citealt{Casey2013}
as COSMOS 450.16 and 450.28). Using ALMA, \citet{Bussmann2015} detected five component SMGs.
We observed component c with MOSFIRE using the same setup and reduced the data in the same manner as \citet{Casey2017}.
To complement the H$\alphaup$-based spectroscopic redshifts, we also consider the CO-based spectroscopic redshifts for three of the component SMGs obtained in other work.

\begin{figure}
\centering
\includegraphics[width=\columnwidth]{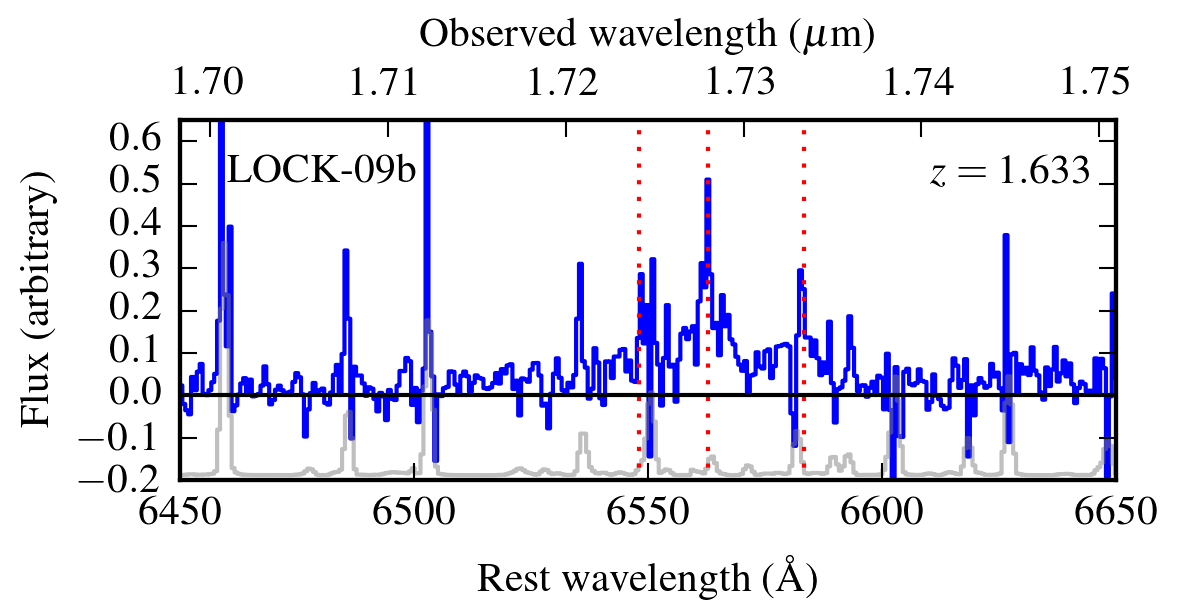}\\
\medskip
\includegraphics[width=\columnwidth]{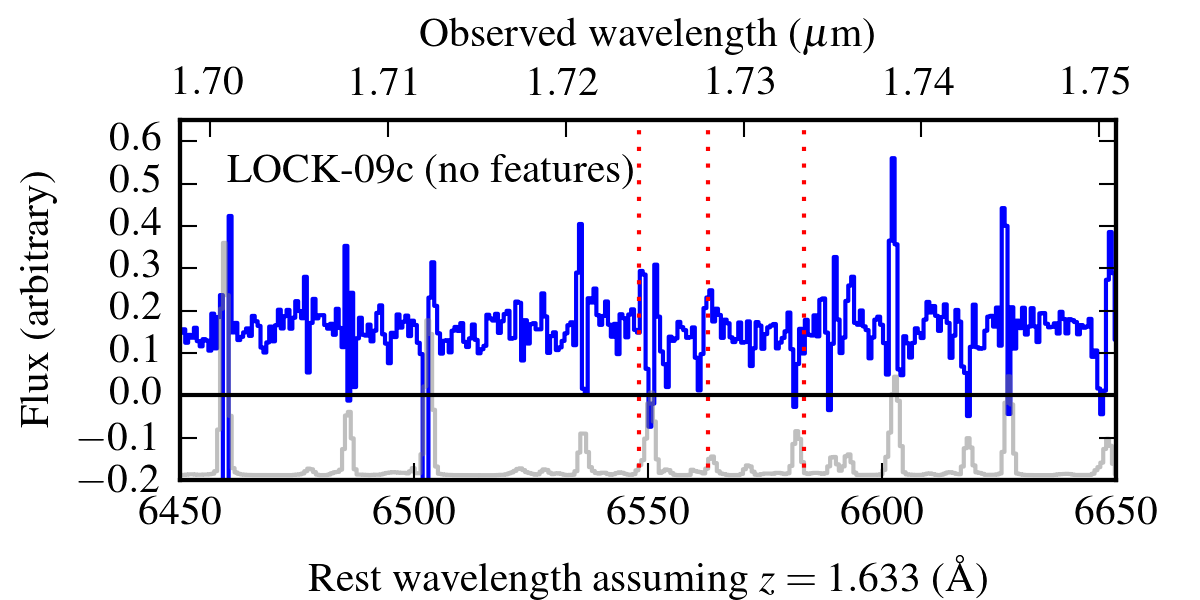}
\caption{Sub-regions of the MOSFIRE \emph{H}-band spectra of the NIR counterparts of LOCK-09b (\emph{top}) and LOCK-09c (\emph{bottom}).
The grey lines show sky spectra (arbitrarily normalised and offset) from \url{https://www2.keck.hawaii.edu/inst/mosfire/sky_lines.html}.
The lines visible in the spectrum of LOCK-09b (marked with vertical dotted red lines) are identified as H$\alphaup$ and the [\ion{N}{ii}] doublet, yielding a firm spectroscopic
redshift of 1.633. Strong continuum emission is detected from LOCK-09c, but H$\alphaup$ emission (at $z = 1.633$) is absent.
Although it is possible that the two component SMGs are at the same redshift, the most parsimonious interpretation
is that the two SMGs are physically unassociated, as also suggested by LOCK-09c's photometric redshift of $0.90 \pm 0.05$;
see text for details.
}
\label{fig:lh09_spectra}
\end{figure}

\subsection{Single-dish submm sources drawn from the literature}

We also analyse four single-dish submm sources from the literature, which, to the best of our knowledge, are the only other suitable multi-component single-dish submm sources
for which rest-frame UV/optical spectra are available for at least two of the component SMGs.
The first two single-dish sources are GOODS 850-13 and 850-15; spectroscopic redshifts for some of the individual component SMGs that comprise these multi-component single-dish sources
were obtained in \citet{Swinbank2004}, \citet{Wang:2011}, and \citet{Barger:2012}.
The remaining two, ALESS~084 and 088, are from \citet{Danielson2017}, who obtained spectra of many of the resolved ALESS SMGs
\citep{Hodge2013}. Although taken at face value, the work of \citet{Danielson2017} contains seven multi-component single-dish submm sources that meet our criteria
(ALESS~017, 041, 067, 075, 080, 084, and 088), we include only two of
them, ALESS~084 and 088, for the following reasons: we exclude ALESS~017, 075, and 080 because for each of these single-dish sources, one of the two component SMGs with spectroscopic redshifts is in
the `supplementary' sample of \citet{Hodge2013} owing to it lying outside the primary ALMA beam, and subsequent observations suggest that most of the `supplementary' sources are artefacts of
poor coverage of the u-v plane (I.~Smail,
private communication). We exclude ALESS~041 because one of its two component SMGs has a spectroscopic redshift and the other's spectrum exhibits continuum without features, but no photometric redshift is available for
the second component. We exclude ALESS~067 because \citet{Danielson2017} claim that this single-dish submm source's two component SMGs are at the same redshift based on morphology,
but they did not obtain two independent spectroscopic redshifts.

Furthermore, some of the component SMGs of the ALESS single-dish submm sources we do include, ALESS~084 and 088, have spectra of marginal quality:
$Q = 3$, indicating redshifts that should not be considered fully independent of the photometric redshifts (and no spectroscopic redshifts based on $Q = 3$
spectra from \citealt{Danielson2017} have been subsequently confirmed; M.~Swinbank, private communication), or $Q = 4$,
corresponding to spectra that exhibit continuum but no sufficiently trustworthy features to assign a spectroscopic redshift (see \citealt{Danielson2017} for further details).
We do not include these components in the `unambiguous sample', which is defined below.

\subsection{Sub-sample definitions}

Below, we analyse the `unambiguous' sub-sample of single-dish submm sources, for which robust spectroscopic redshifts are available for at least two SMGs comprising a given
single-dish submm source, separately. For the full sample, we additionally include component SMGs with less-robust spectroscopic redshifts (specifically, those assigned
$Q = 3$ by \citealt{Danielson2017}) and component SMGs whose spectra exhibit continuum but no features; for the latter, we employ
photometric redshifts to compute the redshift separations of the components. The single-dish submm sources and component SMGs included in the
two sub-samples (with the individual components specified in parentheses) are as follows:
\begin{itemize}
\item `Unambiguous sub-sample': LOCK-03 (components a and b), LOCK-08 (a and b), UDS292 (0 and 1), COSBO3 (a-e),
GOODS~850-13 (b and c), GOODS~850-15 (a and b), and ALESS~088 (1 and 5).
\item `Full sample': all sources/component SMGs in the unambiguous sub-sample, plus LOCK-09 (b and c), UDS306 (0 and 1)\footnote{\label{foot306}
As noted in \tref{tab:sample_details}, the large offset between the potential NIR counterpart to UDS306.2 and the ALMA source
precludes us from assigning a redshift to UDS306.2. We present the XSHOOTER spectrum of the NIR counterpart
for completeness, but we do not include UDS306.2 in the analysis.}, and ALESS~084 (1 and 2).
We also include two additional components SMGs comprising ALESS~088 components 2 and 11 (components 1 and 5
were already included in the unambiguous sub-sample. Although
\citet{Danielson2017} report spectroscopic redshifts for ALESS~088.2 and 11, they are based on marginal spectra ($Q = 3$), so we opt to exclude these
SMGs from the unambiguous sub-sample.
\end{itemize}

\section{Results} \label{S:results}

\begin{figure}
	\centering
	\includegraphics[width=0.95\columnwidth]{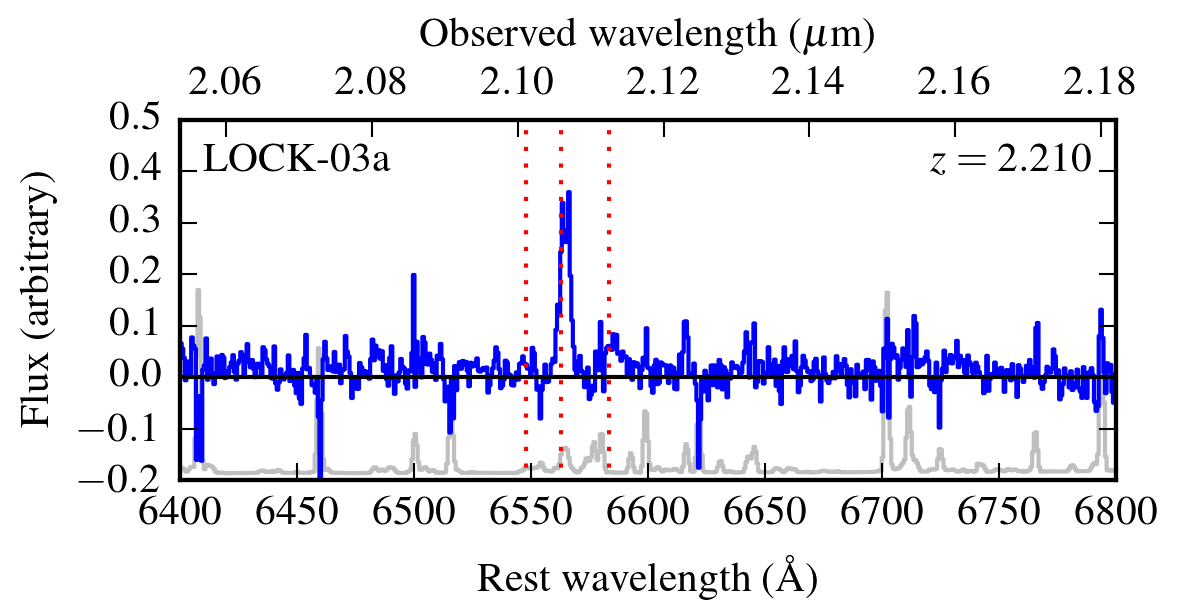}
	\includegraphics[width=0.95\columnwidth]{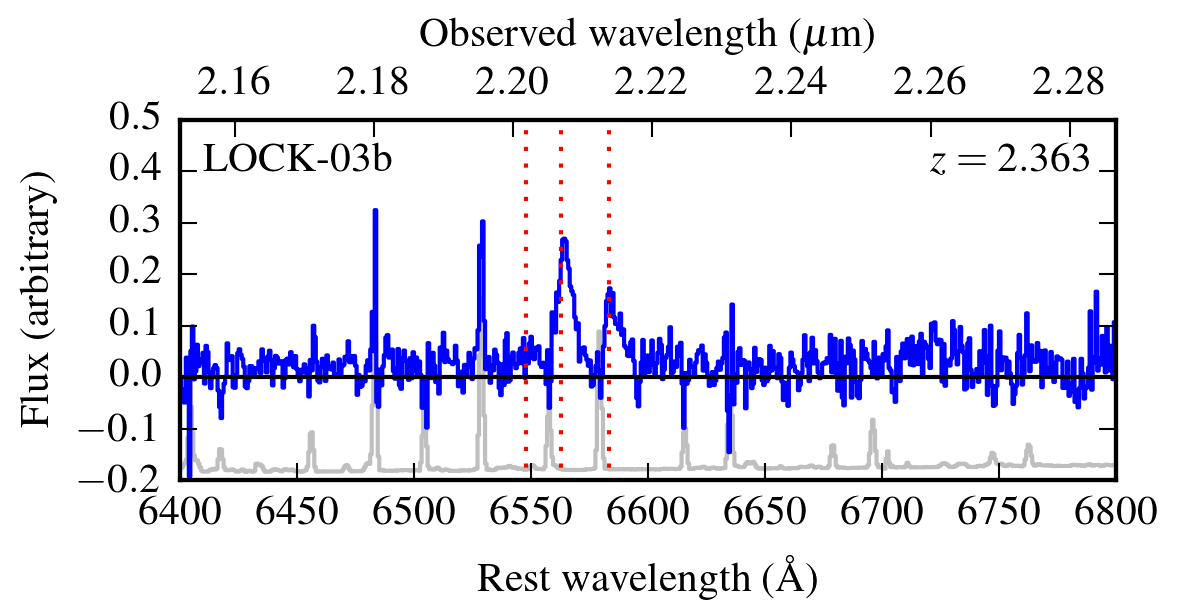}
	\medskip
	\includegraphics[width=0.95\columnwidth]{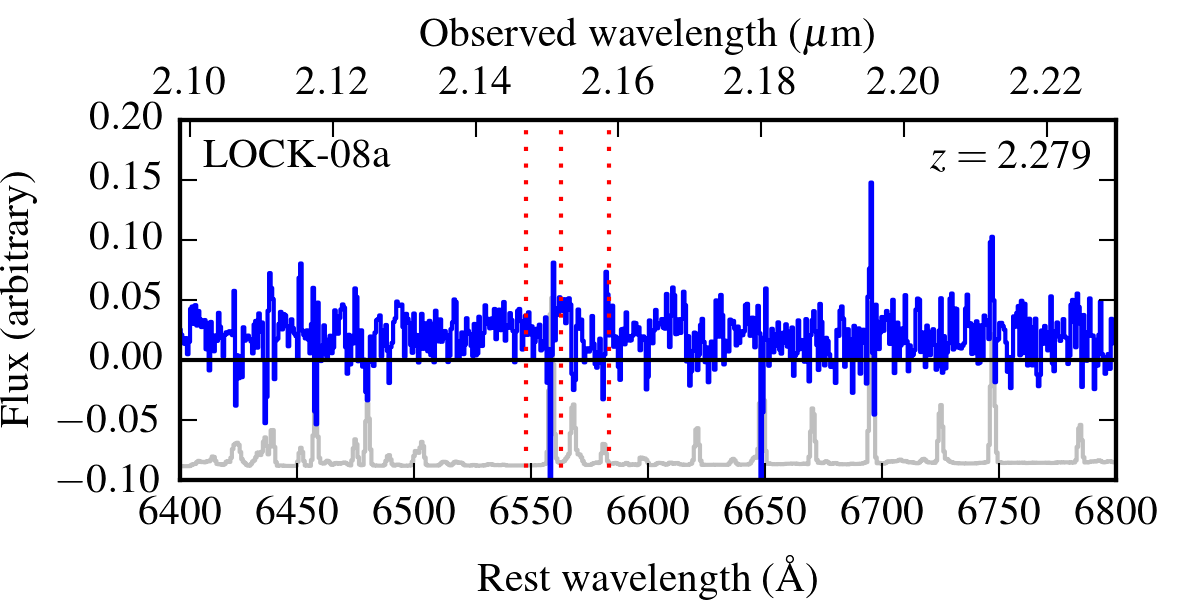}
	\includegraphics[width=0.95\columnwidth]{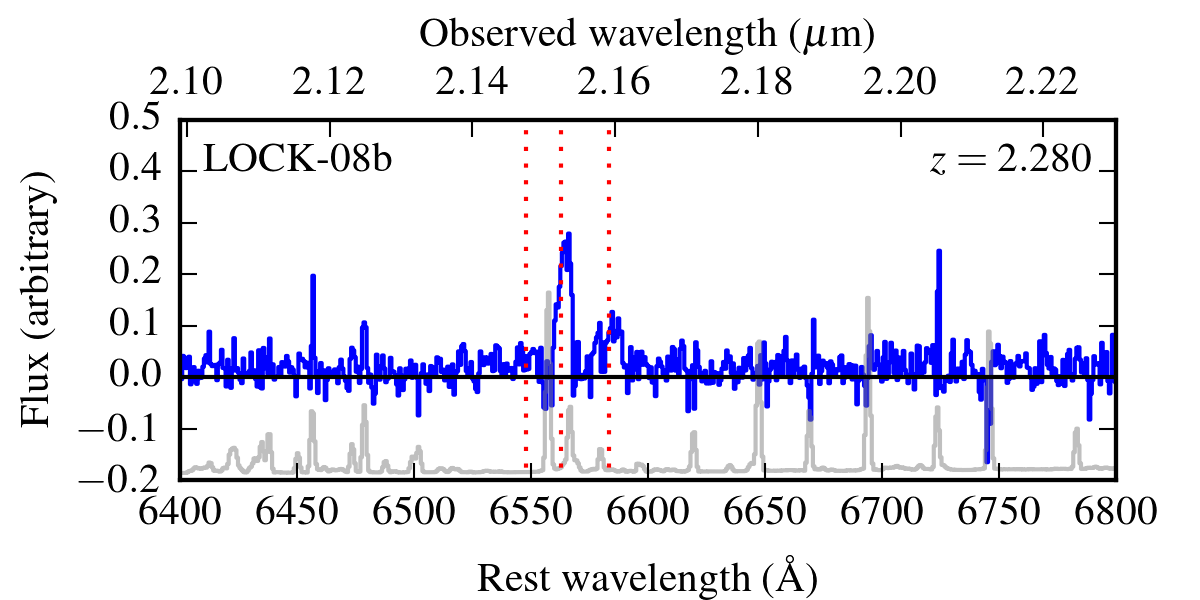}
	\medskip
	\includegraphics[width=0.95\columnwidth]{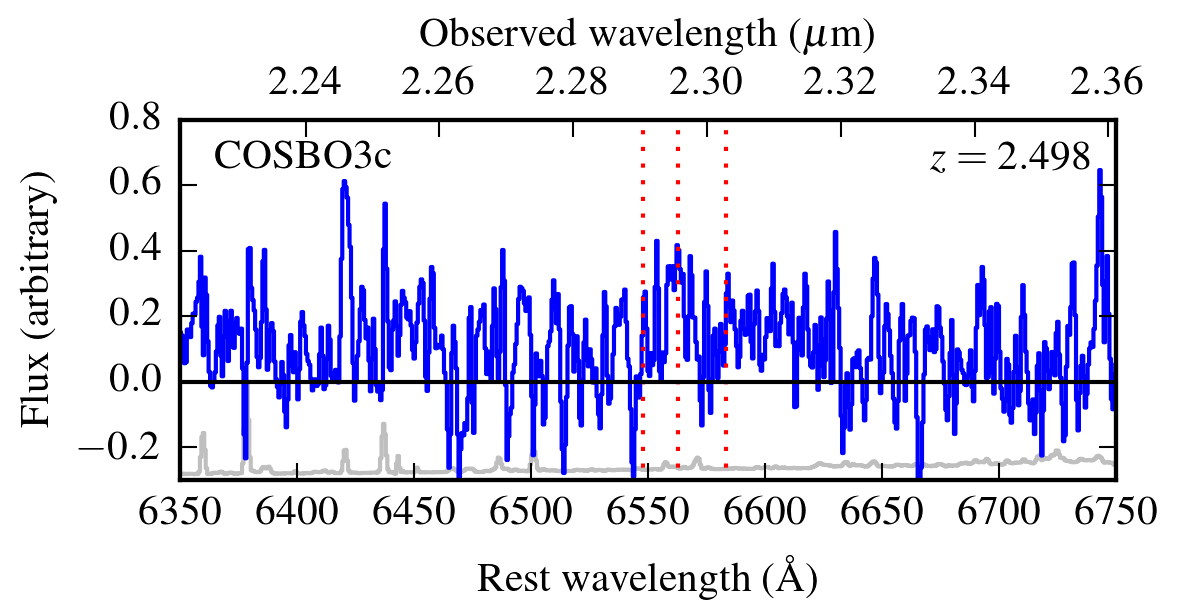} \\
    \caption{Sub-regions of the MOSFIRE spectra of the NIR counterparts of LOCK-03a, LOCK-03b, LOCK-08a, LOCK-08b and COSBO3c near the
    features of interest. The grey lines show sky spectra (arbitrarily normalised and offset) from \url{https://www2.keck.hawaii.edu/inst/mosfire/sky_lines.html}.
    The vertical dotted red lines denote the positions of the H$\alpha$ and the [\ion{N}{ii}] doublet.
    }
    \label{fig:mosfire_spectra}
\end{figure}

\begin{figure}
	\centering
	\includegraphics[width=0.95\columnwidth]{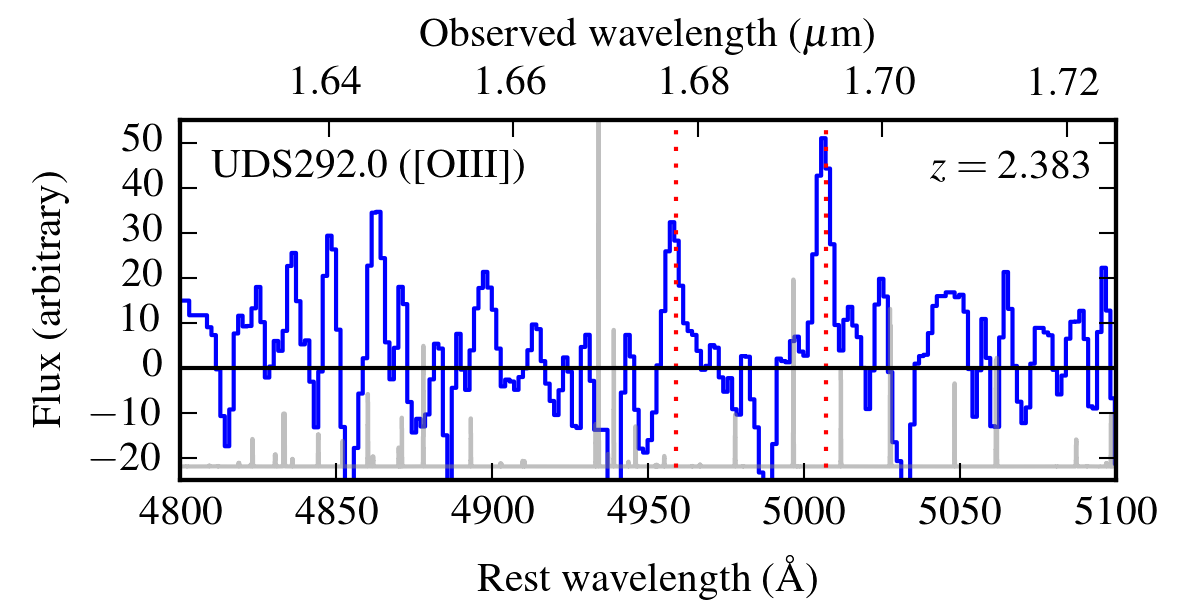}
	\includegraphics[width=0.95\columnwidth]{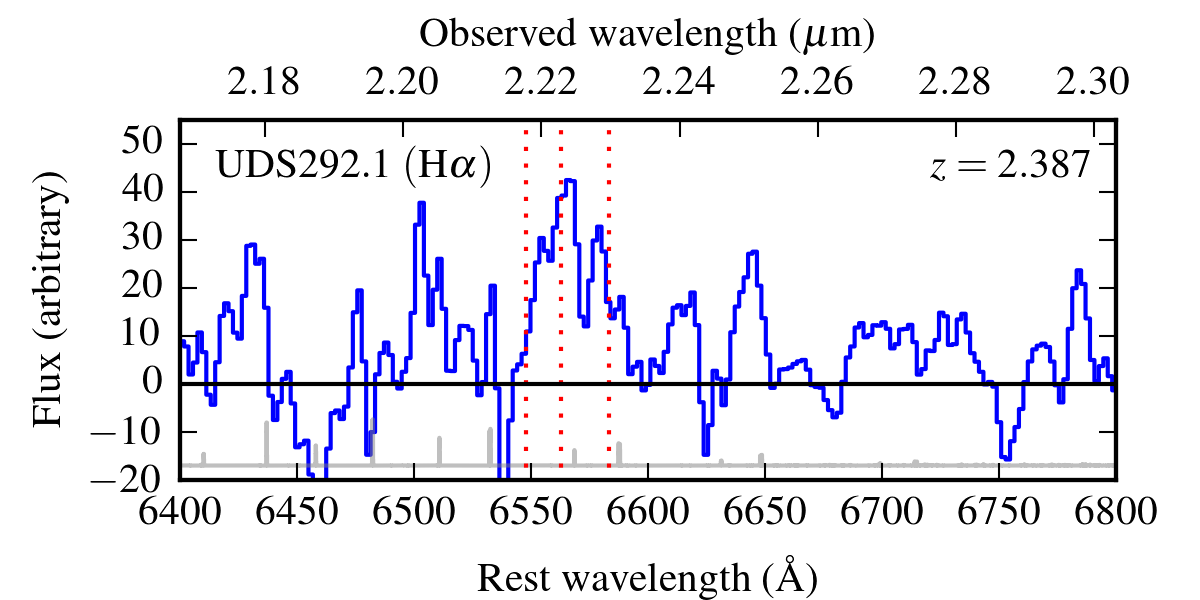}
	\medskip
	\includegraphics[width=0.95\columnwidth]{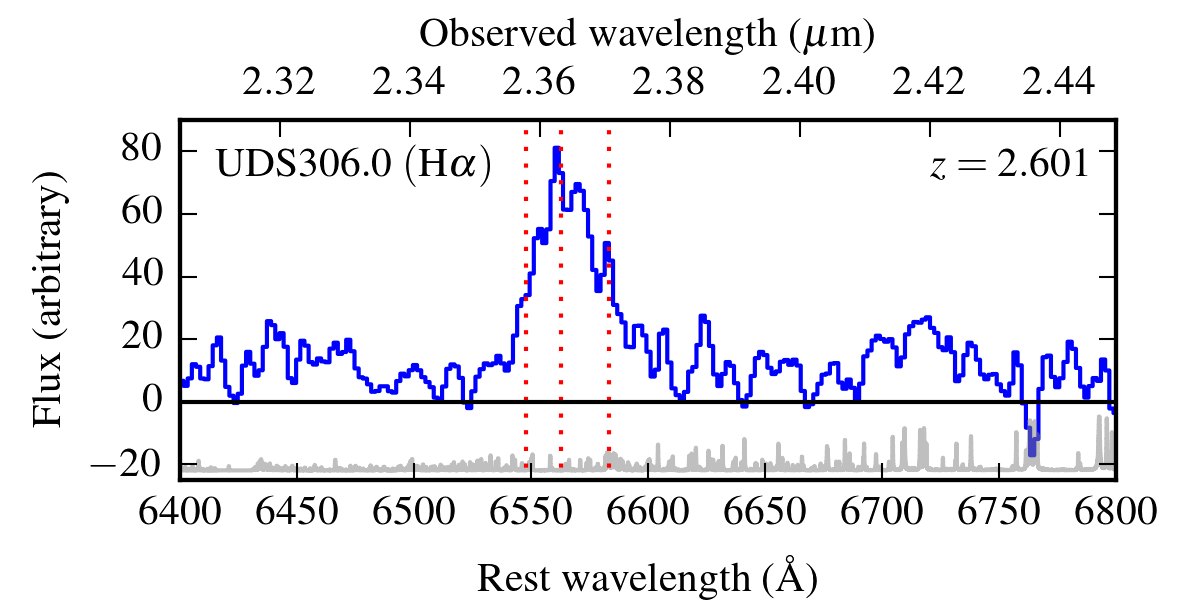}	
	\includegraphics[width=0.95\columnwidth]{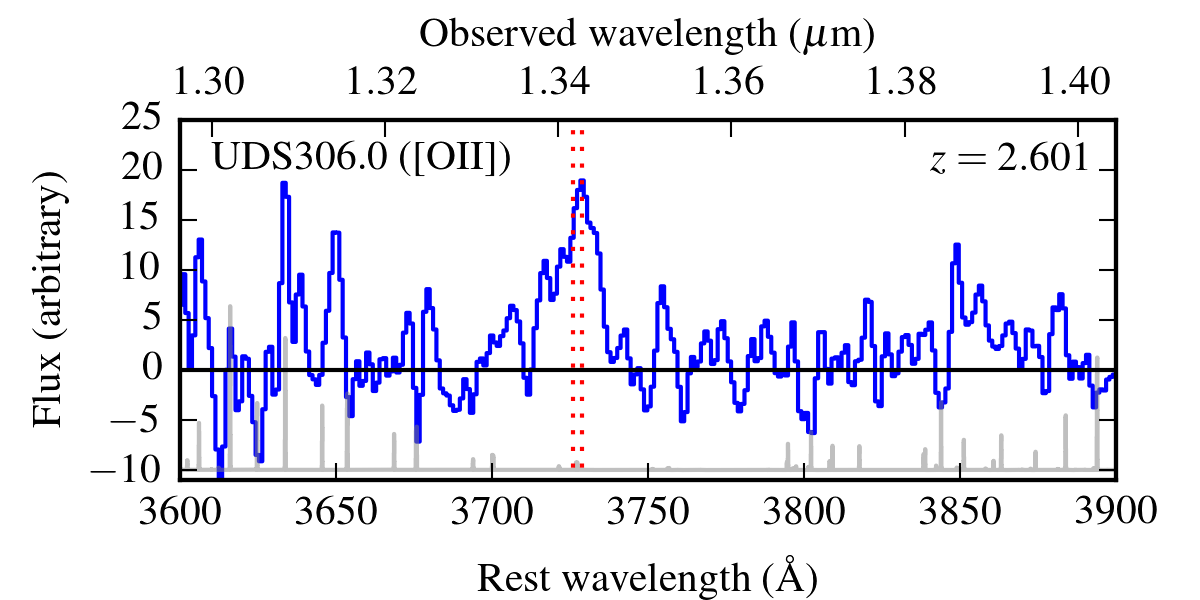}
    \caption{Sub-regions of the GNIRS spectra of the NIR counterparts of UDS292.0, UDS292.1 and UDS306.0 near the features of interest (labeled in the individual
    panels and marked with vertical dotted red lines).
    The grey lines show sky spectra (arbitrarily normalised and offset) obtained from the Gemini Observatory (\citealt{Lord1992};
    \url{http://www.gemini.edu/sciops/telescopes-and-sites/observing-condition-constraints/ir-background-spectra}).
    }
    \label{fig:gnirs_spectra}
\end{figure}

\begin{figure}
	\centering
	\includegraphics[width=0.95\columnwidth]{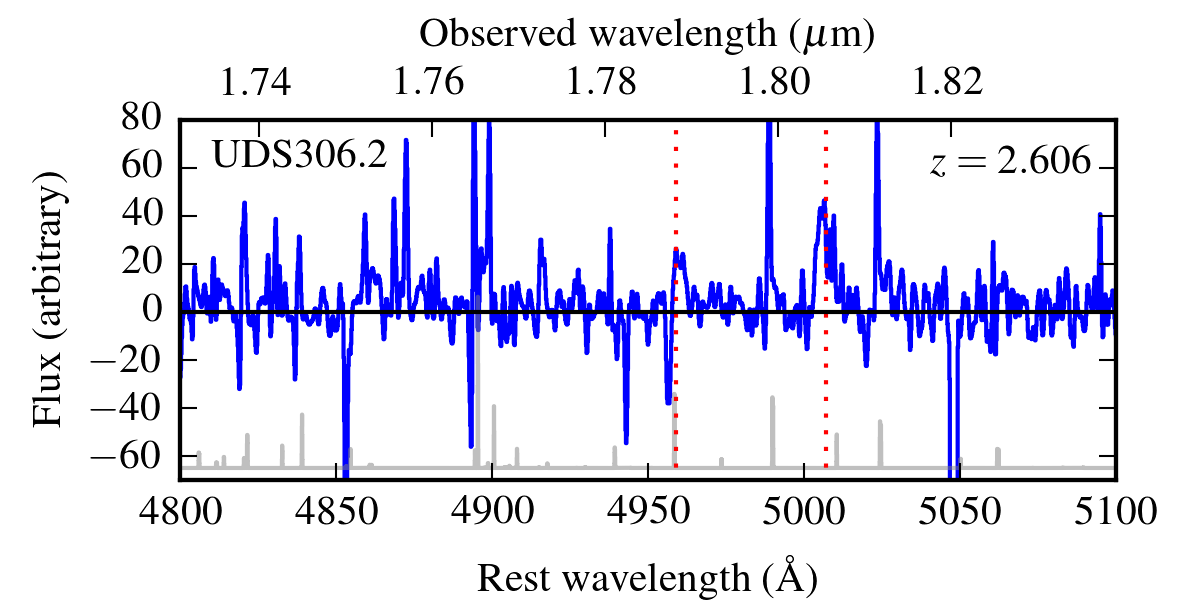}
    \caption{Sub-region of the XSHOOTER spectrum of the potential NIR counterpart of UDS306.2 near the [\ion{O}{iii}] emission lines at 4959 and 5007 \AA ~(marked by the vertical
    dotted red lines). The grey line shows a sky spectrum (arbitrarily normalised and offset) obtained from the Gemini Observatory (\citealt{Lord1992};
    \url{http://www.gemini.edu/sciops/telescopes-and-sites/observing-condition-constraints/ir-background-spectra}).
    }
    \label{fig:xshooter_spectra}
\end{figure}

\subsection{Discussion of individual sources}

\fref{fig:lh09_spectra} shows example MOSFIRE \emph{H}-band spectra of the two component SMGs of one of the single-dish submm sources in our sample,
LOCK-09; the remaining spectra for the components of multi-component single-dish submm sources
obtained in this work using MOSFIRE, GNIRS, and XSHOOTER are shown in Figures \ref{fig:mosfire_spectra}, \ref{fig:gnirs_spectra}, and \ref{fig:xshooter_spectra},
respectively. Thumbnail images showing the interferometric submm sources are available in the works that presented the submm interferometric observations
\citep{Hill2017,Simpson2015b,Simpson2015,Bussmann2015}. Before discussing the contributions of associated and unassociated components,
we first comment on the three single-dish submm sources for which the physical nature of the blending is somewhat ambiguous.

H$\alphaup$ and the [\ion{N}{ii}] doublet are detected in the spectrum of LOCK-09b (top panel of \fref{fig:lh09_spectra}), yielding a firm spectroscopic redshift of 1.633.
The spectrum of LOCK-09c (bottom panel) exhibits strong continuum emission but no obvious features.
We argue that LOCK-09b and LOCK-09c are unlikely to be at the same redshift for the following reasons:
LOCK-09b has an H$\alphaup$ equivalent width (EW) of 120 \AA; if LOCK-09c is at the same redshift,
its H$\alphaup$ EW is $\la 1$ \AA. For LOCK-09b, $L_{\text{H}\alphaup} = (5.65 \pm 0.20) \times 10^{-20}$ W m$^{-2}$
(a 27$\sigma$ detection), corresponding to an unobscured SFR of $(5.6 \pm 0.2)~\msunperyr$. For LOCK-09c, assuming
the same $z = 1.633$, then $L_{\text{H}\alphaup} < 5 \times 10^{-21}$ W m$^{-2}$ (3$\sigma$),
corresponding to a 3$\sigma$ upper limit on the unobscured SFR of $0.5~\msunperyr$.
It is possible that these component SMGs are at the same redshift and that the nebular emission from LOCK-09c is too dust-obscured
to be detected; however, this would require that LOCK-09c has an extremely low H$\alphaup$ equivalent width.
For example, using a sample of 73 local (U)LIRGs, \citet{Poggianti2000} obtained a minimum H$\alphaup$+[\ion{N}{ii}]
EW of 17.7 \AA.
Moreover, given its submm flux density of $S_{850} \approx 5$ mJy,
LOCK-09c likely has an SFR in excess of 500~$\msunperyr$ \citep[e.g.][]{daCunha2015,Cowie2017}; thus, in the scenario in which
this component SMG is at $z = 1.633$ and the H$\alphaup$ emission is simply too (differentially) dust-obscured to be detected,
SFR$_{\text{H}\alphaup}$/SFR$_{\text{IR}} \lesssim 10^{-3}$ (i.e. $A_{\text{H}\alphaup} > 7.5)$.
Such a ratio would be low even for SMGs \citep[e.g.][]{Swinbank2004,Flores2004}, but spectroscopic follow-up studies are likely
biased against SMGs with the highest $A_{\text{H}\alphaup}$ values, and this scenario $(A_{\text{H}\alphaup} > 7.5)$ is not impossible on
these grounds alone. However, given the detection of strong continuum emission, we consider the chance projection scenario
to be more likely.
Finally, LOCK-09c has $\zphot = 0.90 \pm 0.05$; this differs from $z = 1.633$ by $14\sigma$, which further supports our interpretation that LOCK-09c is not at $z = 1.633$
(the photometric redshift of LOCK-09b, $\zphot = 1.46 \pm 0.19$, is consistent with its spectroscopic redshift).
We thus conclude that the most likely explanation for the detection of continuum only from LOCK-09c is that it is a chance projection,
but obtaining a spectroscopic redshift for LOCK-09b is necessary to definitively confirm this conclusion.

For UDS306.1, we detect continuum but no obvious features. However, we do not interpret this component as a chance projection
because its photometric redshift is discrepant from the spectroscopic redshift of UDS306.0 by $<2.5\sigma$. Moreover, the upper limit on the unobscured SFR
($\la 63\, \msunperyr$ at $3\sigma$) is weaker than that for LOCK-09c, so we cannot appeal to the arguments we used above.

The sample drawn from the literature also
contains one single-dish submm source for which continuum only is detected from one component SMG (ALESS~084).
Because H$\alphaup$ luminosities are not reported in \citet{Danielson2017}, we cannot make SFR-based arguments similar
to those made for LOCK-09 above.
For this single-dish source, the photometric redshift of the continuum-only component differs from the spectroscopic redshift of the
other component by $>27\sigma$. However, for the component with a spectroscopic redshift, the photometric redshift
is discrepant from the spectroscopic redshift by almost $23\sigma$, likely because the photometric redshift uncertainties quoted by
\citet{Simpson2014} are unrealistically small, but also perhaps because the spectroscopic redshift's based on marginal-quality
($Q = 3$) spectra from \citet{Danielson2017} are not trustworthy. We thus label this single-dish source as a `likely projection',
but future observations may reveal that the two component SMGs are actually associated.
Regardless, we note that excluding ALESS~084 does not change our qualitative conclusions.

\subsection{Statistical constraints}

\begin{figure}
    \centering
    \includegraphics[width=\columnwidth]{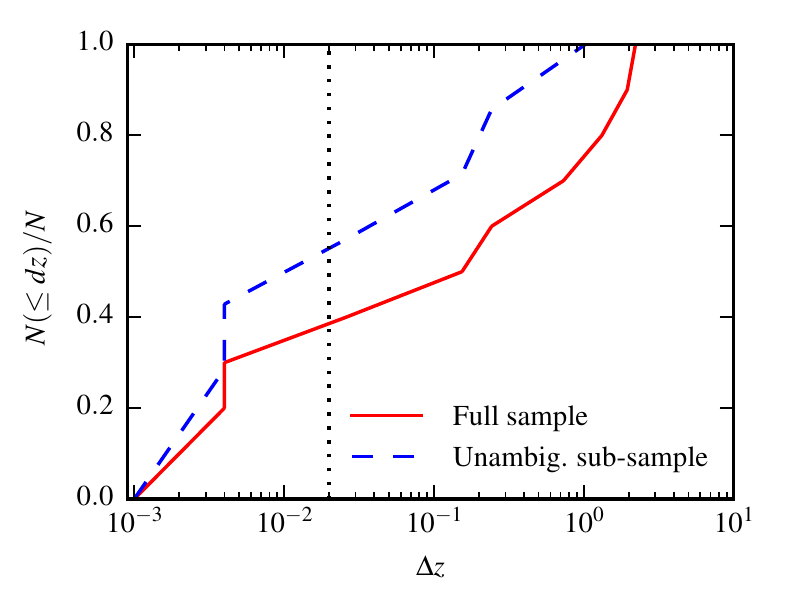}
    \caption{Empirical cumulative distribution functions (CDFs) of $\Deltaup z$ for the multi-component single-dish submm sources in our sample (including those drawn
    from the literature). The \emph{blue dashed} line represents the CDF for the unambiguous sample (i.e. when only robust spectroscopic redshifts are used), whereas
    the \emph{solid red} line shows the CDF for the full sample (for components whose spectra exhibit continuum only, photometric redshifts
    are used to compute $\Deltaup z$).
    The \emph{vertical dotted} line represents the separation between associated and unassociated single-dish submm sources $\Delta z = 0.02$;
    this value is motivated by the bimodal $\Deltaup z$ distributions predicted by multiple theoretical models \citep{HB13,Cowley2015a,MunozArancibia2015}.
    In both sub-samples, of order half of the single-dish submm sources contain at least one unassociated SMG.}
    \label{fig:dz_cdf}
\end{figure}

\ctable[
	caption = {Redshift and angular separations of the components and nature of the multiplicity of single-dish submm sources.\label{tab:all_smgs}},
	center,
	doinside=\small,
	star,
	notespar
]{lcccc}{
	\tnote[a]{Redshift separation of components (see text for definition). Values in parentheses are based on one or more photometric redshifts
	(used only for component SMGs whose spectra exhibit continuum but no features; for these, error bars are quoted for the $\Deltaup z$ values)
	or less-robust ($Q = 3$) spectroscopic redshifts from \citet{Danielson2017}.}
	\tnote[b]{Angular separation of the components used to compute $\Deltaup z$; when there are more than two component SMGs, the pairwise angular separations
	are added in quadrature in the same manner as for $\Deltaup z$. For reference, in the redshift range of interest, $z \sim 1-4$, 1$^{\prime\prime}$ corresponds to 7-8 kpc.}
	\tnote[\dag]{Although we classify COSBO3 as a chance projection because it has $\Deltaup z = 0.026 > 0.02$, we note that this $\Deltaup z$ value is very close
	to the adopted threshold, and this source has been interpreted as a `proto-cluster' by other authors \citep{Casey2015,Wang2016}.}
	}{
																												\FL
	Single-dish ID				&	$\Deltaup z$\tmark[a]			&	Angular separation\tmark[b]	&	Nature of multiplicity		\NN
							&								&	(arcsec)					&						\ML
	LOCK-03					&	0.154						&	11.3						&	unassociated			\NN
\noalign{\smallskip}
	LOCK-08					&	0.001						&	24.7						&	associated			\NN
\noalign{\smallskip}
	LOCK-09					&	($0.73\pm0.05$)				&	(12.0)					&	unassociated			\NN
\noalign{\smallskip}
	UDS292					&	0.004						&	6.7						&	associated			\NN
\noalign{\smallskip}
	UDS306					&	$\left(1.32^{+0.06}_{-0.53}\right)$	&	(6.1)						&	ambiguous			\NN
\noalign{\smallskip}
	COSBO3\tmark[\dag]		&	0.026						&	18.1						&	unassociated			\NN
\noalign{\smallskip}	
	GOODS 850-13			&	0.243 						&	16.4						&	unassociated			\NN
\noalign{\smallskip}
	GOODS 850-15			&	0.004						&	2.0						&	associated			\NN
\noalign{\smallskip}
	ALESS 084				&	$\left(2.22^{+0.19}_{-0.08}\right)$	&	8.7						&	unassociated			\NN
\noalign{\smallskip}
	ALESS 088				&	1.026 (1.951)					&	15.0 (17.9)				&	unassociated			\LL
}

For each of the multi-component single-dish submm sources in the unambiguous sample, we compute the difference in redshift between the components for which
we have spectroscopic redshifts, $\Deltaup z$; for single-dish sources with more than two component SMGs, following \citet{HB13}, we compute the redshift
separation of each subdominant component from the brightest component and add these separations in quadrature.
For the full sample, we compute $\Deltaup z$ using spectroscopic redshifts when available and photometric redshifts only for component SMGs without spectroscopic
redshifts.\footnote{The $\Deltaup z$ values based on photometric redshifts should be interpreted with caution because obtaining accurate photometric redshifts 
for dust-obscured galaxies is challenging. For many SMGs, the spectroscopic redshifts are formally significantly discrepant from the photometric redshifts reported in other works.
These discrepancies suggest that the original works underestimated the uncertainties on their photometric redshifts.
However, even if the uncertainties on the photometric redshifts were multiplied by a factor of a few to account for this likely underestimation,
the $\Deltaup z$ values computed using photometric redshifts for LOCK-09, ALESS~084, and ALESS~088 would remain significantly
greater than 0.02, and thus their classification as chance projections should be robust to this issue.}
We then classify the physical nature of the multiplicity of each single-dish submm source using the criterion defined in \citet{HB13},
which was selected because this value separates the two peaks of the bimodal $\Deltaup z$ distribution predicted by their theoretical model
(see their fig. 4):
associated sources are those with $\Deltaup z \le 0.02$, whereas those with greater $\Deltaup z$ values are considered chance projections of at least
one unassociated component SMG. Na\"ively, considering only the velocity difference between two galaxies and asking whether the galaxies
are bound, a threshold of $\Deltaup v \ga 1000$ km s$^{-1}$, or $\Deltaup z \ga 0.003$, would be sufficient to identify unbound -- and thus non-merging -- pairs,
and an even lower threshold could likely be used. However, we also wish to classify unbound but still associated pairs, such as those contained in the same
`proto-cluster' or dark matter filament, as physically associated. Additionally, for single-dish sources with more than two component SMGs, the fact that we add the pairwise
redshift separations in quadrature means that the velocity difference inferred from the $\Deltaup z$ value will be greater than the velocity differences between
individual galaxies. 
Moreover, because the distribution of $\Deltaup z$ in predicted by theoretical models \citep{HB13,Cowley2015a,MunozArancibia2015} is strongly bimodal,
using a somewhat smaller or larger threshold does not affect the model predictions. However, one of our single-dish sources,
COSBO3, has a $\Deltaup z$ value very close to this threshold, $\Deltaup z = 0.026$. Although we formally treat this as a chance projection
to ensure the fairest possible comparison with the model predictions, we note that this classification is sensitive to the exact threshold employed,
and this association of SMGs has been previously interpreted as a `proto-cluster' \citep{Casey2015,Wang2016}.

In \fref{fig:dz_cdf}, we show cumulative distribution functions (CDFs) of $\Deltaup z$ for the two sub-samples defined above;
the redshift separations and nature of the component multiplicity are presented in \tref{tab:all_smgs}.
The dashed blue line represents the CDF of the unambiguous sample (i.e. only robust spectroscopic redshifts are used).
The solid red line shows the CDF of the full sample.
The CDFs reveal that for both sub-samples, of order half the single-dish submm sources are chance projections of at least one unassociated component (i.e.~
have $\Deltaup z > 0.02$).
For the unambiguous sub-sample and the full sample, the mean $\Deltaup z$ values are $0.21 \pm 0.05$ and $0.67 \pm 0.23$, respectively.

Of the seven single-dish submm sources in the unambiguous sub-sample, in three sources $\left(3/7, \text{ or } 43\substack{+39 \\ -33} \text{ per cent}
\right)$\footnote{Throughout the work, the quoted uncertainties on percentages correspond to
95-per cent binomial confidence intervals \citep{ClopperPearson1934}.} --
LOCK-08, UDS292, and GOODS~850-15 -- the two component SMGs for which spectroscopic redshifts are available are physically associated.
Interestingly, because the components of LOCK-08 have a projected separation of $\sim 25$ arcsec
($\sim 200$ kpc), they are unlikely to be undergoing a merger but rather simply part of the same dark matter filament.
The remaining four single-dish sources $\left(4/7, \text{ or } 57\substack{+33 \\ -39} \text{ per cent} \right)$ in the unambiguous sub-sample (LOCK-03, COSBO3,
GOODS~850-13, and ALESS~088) are classified as chance projections of at least one unassociated SMG, although as already noted above,
COSBO3 has a $\Deltaup z$ value only slightly greater than the threshold for a chance projection; given the significant
uncertainties in the associated and unassociated fractions as a result of the small sample size, classifying COSBO3
as physically associated or removing it entirely would not materially affect our conclusions. LOCK-03 is particularly
interesting because the MOSFIRE spectrum of one of the resolved SMGs, LOCK-03a, exhibits two kinematically
distinct components separated by $0\farcs8$ ($\sim 7$ kpc) and 160 km s$^{-1}$. Thus, it appears that the ALMA source LOCK-03a
corresponds to a late-stage, pre-coalescence merger (i.e.~is a `galaxy-pair SMG' in the parlance of \citealt{H11}),
and it may be resolved into two separate sources in future higher-resolution interferometric submm observations.
The other ALMA component, LOCK-03b, which has a submm flux density equal to that of LOCK-03a, is at a significantly
different redshift (the two sources are separated by $\Deltaup z = 0.154$). LOCK-03 thus appears to be a blend of an ongoing merger
and an unassociated SMG.

Turning to the full sample (i.e.~using photometric redshifts to compute $\Deltaup z$ for component SMGs from which continuum only was detected), an additional three single-dish submm sources
(LOCK-09, UDS306 and ALESS~084) would na\"ively be classified as chance projections based on their having $\Deltaup z > 0.02$.
However, regarding UDS306, because the photometric redshifts of the two component SMGs are discrepant by $<2.5\sigmaup$, we consider the nature of the blending
unconstrained. Thus, of the nine single-dish sources in the full sample for which we have firm or tentative constraints on the nature of the multiplicity,
six $\left(6/9, \text{ or } 67\substack{+26 \\ -37} \text{ per cent}\right)$ are comprised of at least one unassociated component SMG.
The full sample thus contains a greater fraction of chance projections than the unambiguous sub-sample, but given the large uncertainties, the difference is
not statistically significant.

\section{Summary and discussion} \label{S:sum}

We have obtained spectroscopic constraints on the redshifts of individual components of six multi-component single-dish submm sources
to investigate the nature of single-dish submm source multiplicity (i.e.~whether the component SMGs are physically associated or chance projections).
We supplemented our sample with four single-dish sources from the literature. Of the seven multi-component single-dish submm sources in our sample for which robust spectroscopic redshifts
are available for at least two component SMGs, only three $\left(3/7, \text{ or } 43\substack{+39 \\ -33} \text{ per cent} \right)$ are clearly blends of physically associated
(but not necessarily merging) SMGs.
Considering also constraints based on detection of continuum but no features, for which photometric redshifts are used to calculate the redshift separation,
and spectroscopic redshifts based on marginal spectra, the nature of the multiplicity of nine of the single-dish sources can be tentatively constrained.
Of these, six $\left(6/9, \text{ or } 67\substack{+26 \\ -37} \text{ per cent}\right)$ are comprised of at least one unassociated component SMG.
The nature of the multiplicity of one single-dish source, UDS 306, is unconstrained by the available data.

This work is the first to place statistical constraints on the relative contributions of physically associated SMGs
and chance projections to the multi-component single-dish submm source population. Such constraints can help distinguish amongst competing theoretical
models for the SMG population. For example, in models in which mergers dominate the submm counts, the components of multi-component
SMGs should correspond to pre-coalescence mergers \citep{H11,H12,HN13} and thus be close in terms of both redshift and angular separation.
However, if late-stage merger-induced starbursts dominate the single-dish submm source population, high multiplicity would not be expected because
such sources would not be resolved into multiple component SMGs.
At the least, the fact that we observe a significant fraction of chance projections -- in addition to single-dish submm sources comprised of
associated SMGs that are widely separated -- demonstrates that the classical view of bright submm
sources as predominantly merger-induced starbursts \citep[e.g.][]{Engel:2010} is incomplete. Put otherwise, the combined effect of mergers
-- i.e. merger-induced starbursts and blending of pre-coalescence merging galaxies can both result in bright single-dish submm sources \citep{HN13}
-- alone is insufficient to explain the observed submm counts if chance projections of unassociated SMGs are common amongst the single-dish
submm source population, as our results suggest.

Although few theoretical models of single-dish submm sources have
treated blending of physically unassociated SMGs, let alone chance projections, both \citet{HB13} and
\citet{Cowley2015a} concluded that the majority of multi-component single-dish submm sources should be comprised
of at least one physically unassociated SMG. However, the quantitative predictions of these models differ:
whereas \citet{Cowley2015a} predict that an almost-negligible fraction of multi-component single-dish submm sources are comprised of solely physically
associated SMGs (see their figure 8),
in the \citet{HB13} model, such sources account for $\sim15$ per cent of the population. 
The fraction of single-dish sources comprised of at least one unassociated component SMG found in this work ($\mathbf{57\substack{+33 \\ -39}}$
and $67\substack{+26 \\ -37}$ per cent for the unambiguous sub-sample
and the full sample, respectively) is more consistent with the predictions of \citet{HB13} than those of \citet{Cowley2015a}.
However, the model predictions are sensitive to details such as the detection limits of the single-dish and interferometric observations
considered. Moreover, the sample here may be subject to various biases and is modest in size.
For these reasons, we caution against over-interpreting this comparison, and we defer a detailed comparison with models
to future work. Nevertheless, our results qualitatively support the predictions of the aforementioned theoretical works, and future studies of
larger samples should help distinguish amongst these and other theoretical models intended to reproduce the submm number counts.

We end with some caveats. In addition to being modest in size, the present sample is likely to be biased. Regarding our own observations, five of the six
single-dish sources were selected from amongst
the brightest SMGs in the S2CLS, and it is expected that the contribution of chance projections depends on the single-dish flux \citep{HB13,Cowley2015a}.
Moreover, our SMA data do not have sufficient resolution to distinguish
mergers near coalescence (with projected separation $\lesssim 2$ arcsec, or $\la 15$ kpc at the relevant redshifts), which implies a
bias against physically associated multiples; the ALMA data are affected by the same bias but to a lesser extent. Regarding the sample
drawn from the literature, no uniform selection was applied, and data censoring is likely to be an issue.
Despite the above
caveats, our work demonstrates that the contribution of chance projections to the single-dish submm source population cannot be
ignored and provides motivation for constraining the physical nature of single-dish submm source multiplicity using a significantly larger,
uniformly selected sample in future work.

\section*{Acknowledgements}

CCH thanks Nick Scoville for encouragement to try his hand at observing,
Ian Smail for inspiration via his Fifth Rule of Observational Cosmology
(``If you see an observational paper with a theorist as lead author be \textit{very} afraid''),
and Allison Strom for advice when planning the observations. We
also thank James Simpson and Ian Smail for comments on an earlier version of
the manuscript, Mark Swinbank for useful discussion, and the anonymous
referee for useful comments that led us to improve the manuscript.
The Flatiron Institute is supported by the Simons Foundation.
Some of this work was supported by a NASA Keck PI Data Award, administered by the NASA Exoplanet Science Institute.
EI acknowledges partial support from FONDECYT through grant no.\,1171710.
MJM acknowledges the support of the National Science Centre, Poland through POLONEZ grant 2015/19/P/ST9/04010.
This project has received funding from the European Union's Horizon 2020 research and innovation programme under Marie
Sk{\l}odowska-Curie grant agreement no.\,665778.

Some of the data
presented herein were obtained at the W.~M.~Keck Observatory, which is operated as a scientific partnership among the California Institute of Technology,
the University of California and the National Aeronautics and Space Administration. The Observatory was made possible by the generous financial
support of the W.~M.~Keck Foundation.
The authors wish to recognize and acknowledge the very significant cultural role and reverence that the summit of Mauna Kea has always had
within the indigenous Hawaiian community.  We are most fortunate to have the opportunity to conduct observations from this mountain.
Partially based on observations obtained at the Gemini Observatory, which is operated by the Association of
Universities for Research in Astronomy, Inc., under a cooperative agreement with the NSF on behalf of the Gemini partnership: the National
Science Foundation (United States), the National Research Council (Canada), CONICYT (Chile), Ministerio de Ciencia, Tecnolog\'{i}a e
Innovaci\'{o}n Productiva (Argentina), and Minist\'{e}rio da Ci\^{e}ncia, Tecnologia e Inova\c{c}\~{a}o (Brazil).

This research has made use of the NASA/IPAC Extragalactic Database (NED), which is operated by the Jet Propulsion Laboratory,
California Institute of Technology, under contract with the National Aeronautics and Space Administration,
the VizieR database of astronomical catalogues \citep{Ochsenbein2000},
NASA's Astrophysics Data System Bibliographic Services, the \href{www.arxiv.org}{arXiv.org} preprint server, and
Astropy, a community-developed core Python package for astronomy \citep{astropy}.

\bibliographystyle{mnras}
\bibliography{smg_obs,std_citations,smg}

\bsp
\label{lastpage}

\end{document}